# Unlocking Electro-optic Resonant Phase Shifting for Multi-dimensional, Ultra-dynamic Photonic Switches

Lingzhi Luo, Rui Ma, Richard V. Penty, *Senior Member, IEEE* and Qixiang Cheng

*Abstract*—Optical circuit switching is connection-oriented, being deterministic through the reservation of a complete wavelength channel or spatial path for a certain period. However, this comes at a trade-off against link dynamics, and overall capacity can thus be constrained by the time slot reservations, especially for switches with microsecond- to millisecond-scale reconfiguration times. This situation calls for a new multi-dimensional switching paradigm that fully exploits not only the space and wavelength domains but also with nanosecond-scale reconfigurable capability in the time domain. In this work, we focus on the exploitation of micro-ring resonant phase shifters (RPSs) that are wavelength selective for optical switching in a single plane. By proposing an innovative analytical method with transmission circle chart, we fully unlock the power of RPS with nanosecond-scale reconfigurability and the capability to arbitrarily manipulate its phase and amplitude response. Such a compact model offers fresh insights into designs with under and critically coupled RPSs beyond the commonly explored over-coupling condition. This creates not only versatile switch elements but also perfect absorbers for robust multi-wavelength operations. The proposed device can bring about a breakthrough in the optical switching capacity that potentially addresses the challenges faced by modern data center networks, as well as other photonic circuits for high-throughput signal processing.

## I. INTRODUCTION

THE increasing demands for emerging high-volume and latency-constrained applications, such as artificial intelligence (AI), autonomous vehicles (AV), and augmented/virtual reality (AR/VR), are imposing stringent requirements on data center network capacity, latency, energy consumption, and guaranteed delivery [1], [2]. Current intra-data center interconnects using static node configurations are being pushed to their limits by the diverse demands of this set of applications. Flexible architectures that match their capacity to an application's traffic pattern can deliver optimized connectivity while eliminating over-provisioned resources [3]. Transparent optical switches are the key enablers.

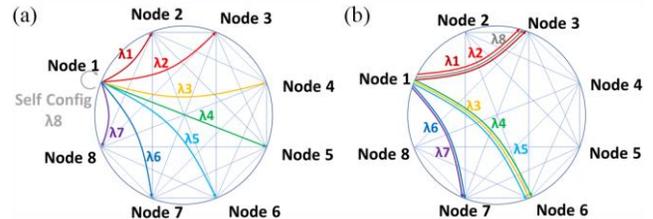

Fig. 1. (a) All-to-all reconfigurable interconnection by allocating a wavelength to any pair of nodes. (b) Bandwidth steering by grouping multiple wavelengths between hot nodes.

Mass-manufacturable photonic integration technologies are favorable for data centers, where cost-effectiveness is a primary concern [4]. Thousands of photonic components have been integrated onto a single chip for large-scale switch fabrics [5], [6]. These demonstrations are broadband and capable of handling wavelength division multiplexing (WDM) data for modulation-agnostic switching, but only reconfigurability in the space domain has been exploited for coarse routing. Switching at the wavelength granularity, which allows for finer-level information access, is becoming vital as short-reach interconnects approach petabit/s-scale capacity with massively parallel wavelength channels [7]. Apart from the fact that only few explorations have been made of chip-scale space-and-wavelength switches, designs often adhere to parallel space switching planes, bookended by wavelength (de)-multiplexers [8], posing immense challenges to the complexity of photonic integrations. In contrast, space-and-wavelength switching can be realized using wavelength-selective elementary switch cells, such as add-drop micro-ring resonators (MRRs) [9], [10]. These critically-coupled MRR filters are often susceptible to fabrication variations and still worse, to electro-absorption losses. This leaves little room for E-O MRR switch fabrics and indeed, this type of device has been sparsely reported [11]. Figure 1 illustrates a multi-dimensional switch that resolves the trade-off between flexibility and determinism. This is achieved by offering reconfigurable interconnects from all-to-all links (Fig. 1(a)) to imbalanced group-to-group traffic through bandwidth steering (Fig. 1(b)). A space-and-wavelength switch fabric forms the core for arbitrarily routing any combination of wavelengths from any input to any output. An extra key lies in

Manuscript received XXXX XX, XXXX; revised XXXX XX, XXXX; accepted XXXX XX, XXXX. This work was supported by European Union's Horizon Europe Research and Innovation Program under Agreement 101070560 (PUNCH).

The authors are with the Centre for Photonic Systems, Electrical Engineering Division, Department of Engineering, University of Cambridge, Cambridge, CB3 0FA, U.K. (e-mail: ll672@cam.ac.uk, rm2126@cam.ac.uk, rvp11@cam.ac.uk, and qc223@cam.ac.uk). Lingzhi Luo and Qixiang Cheng is also with GlitterinTech Limited, Xuzhou, China.

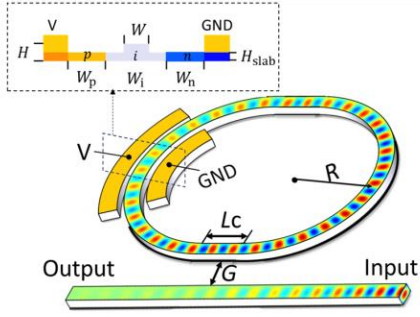

Fig. 2. Schematic diagrams of Electro-optical RPS. *R*: radius of the bend waveguide, *Lc*: the length of straight coupling waveguide, *G*: the gap between the straight coupling waveguide and the bus waveguide., *W*: waveguide width. *H*: waveguide height, $H_{\text{slab}}$:slab thickness. *W*p, *W*i, *W*n: width of p zone, intrinsic zone and n zone.

the ultra-fast reconfigurability, which is a cornerstone that enables unparalleled dynamics by multiplexing in the time domain. Prior studies used MRRs to build space-and-wavelength switches[12], [13]; however, the stringent management on fabrication variations poses a challenge. An alternative approach utilized over-coupled MRRs as phase shifters but with transmission asymmetry and/or high insertion loss[14], [15], [16].

In this paper, by proposing a powerful and innovative analytical method with transmission circle chart, we fully unlock the power of RPSs with electro-optic (E-O) nanosecond-scale reconfigurability. Manipulation of the coupling strength, which is the key to tweaking its performance, can be simply done by adjusting the position of points on the transmission circle chart and altering the circle shape. We extend the model beyond the commonly utilized over-coupling regime to explore under and critically coupled conditions and outline their role, creating both versatile multi-wavelength switch elements (SE) and perfect absorbers. As a benchmark, our exploration in the design of RPSs confirms its superiority over conventional add-drop MRRs. This presents as an efficient design methodology for tailoring RPS elements, potentially bringing about a breakthrough in the field of optical switching within intra-data center networks and other applications requiring ultra-dynamic and ultrafast optical signal processing.

## II. Transmission Circle Chart of RPSs

MRRs exhibit a strong resonance phenomenon, wherein even slight deviation from the resonance results in a significant phase change. This character can be harnessed for efficient narrow-band phase manipulation, which forms RPSs. A powerful analytical model, transmission circle chart, is developed here to precisely and visually describe the behavior of both thermo-optic (T-O) and E-O RPSs.

Figure 2 illustrates the structure of a racetrack E-O RPS with key geometric parameters. Its optical field transmission coefficient ($t_{MRR}$) can be expressed as [17]:

$$t_{MRR}(\phi) = e^{i(\pi+\phi)} \frac{a - re^{-i\phi}}{1 - rae^{i\phi}} \quad (1)$$

where $a$ and $r$ are the single-pass amplitude transmission coefficient and the self-coupling coefficient, respectively. $\phi = \beta L$ is the single-pass phase shift, with $L$ as the round-trip length and $\beta$ as the propagation constant. The argument and the square of modulus of $t_{MRR}$ are defined as the effective phase shift ($\varphi$) and power transmission coefficient ($T$), respectively, which can be given as:

$$\varphi(\phi) = \pi + \phi + \arctan\left(\frac{r\sin(\phi)}{a - r\cos(\phi)}\right)$$
$$+ \arctan\left(\frac{ra\sin(\phi)}{1 - ra\cos(\phi)}\right), a > r$$
$$\varphi(\phi) = -\arctan\left(\frac{a\sin(\phi)}{r - a\cos(\phi)}\right) \quad (2)$$
$$+ \arctan\left(\frac{ar\sin(\phi)}{1 - ar\cos(\phi)}\right), a \leq r$$

$$T(\phi) = |t_{MRR}(\phi)|^2 = \frac{a^2 - 2ra\cos\phi + r^2}{1 - 2ar\cos\phi + (ra)^2} \quad (3)$$

To ascertain the correlation between $\varphi(\phi)$ and $|t_{MRR}(\phi)|$, we decompose $t_{MRR}$ into its real and imaginary components $(x, iy)$ and plot them on the complex plane with varying $\phi$. Consequently, with $\phi$ increasing, $t_{MRR}$ rotates counterclockwise and finally traces out a circular locus that satisfies the equation below:

$$\left(x - \frac{(a^2-1)r}{a^2r^2-1}\right)^2 + y^2 = \left(\frac{a(r^2-1)}{a^2r^2-1}\right)^2 \quad (4)$$

This creates a transmission circle chart that is centered at $\alpha = ((a^2-1)r/(a^2r^2-1), 0)$ with a radius of $R_c = a(r^2-1)/(a^2r^2-1)$. The angle and magnitude of vectors from the origin (point O) to a specific point on the circle correspond to $\varphi(\phi)$ and $|t_{MRR}(\phi)|$, respectively. Vector OA that points from O to the leftmost point on the circle corresponds to $t_{MRR}(\phi)$ at resonance. Based on the relative values of $a$ and $r$, the transmission circle chart can be divided into 3 cases that are illustrated by Fig. 3(a)-(c), respectively. (I) When $a < r$, the RPS is in the under-coupled regime and O is outside the circle. Therefore, there are two extremums ($\pm\theta_m$) on the effective phase shift curve as shown in Fig. 3(d), corresponding to the vectors OB and OC that are tangent to the circle in Fig. 3(a). (II) When $a = r$, the RPS is critically coupled, and O is on the circle so the transmission is zero. When the ring deviates from resonance by a plus or minus infinitesimal value, the angle of the corresponding vector undergoes an abrupt transition to $\pm\pi/2$. This accounts for the phase jump observed in Fig. 3(e). (III) When $a > r$, the RPS is in the over-coupled regime and O is inside the circle so the angle of the vector from O to the point on the circle can take all values from 0 to $2\pi$ continuously, as shown in Fig. 3(f).

When the RPS is implemented with T-O phase controllers, i.e., heaters, $a$ remains constant so analyses can be conducted in a single transmission circle chart with a fixed center of circle and radius. As for E-O phase controllers, i.e., by carrier injections, $a$ would vary due to electro-absorption losses when injecting different carrier concentrations for phase shifting. Consequently, each point of $t_{MRR}$ corresponds to a distinct transmission circle chart, as depicted in Fig. 4(a). As $a$ decreases, $\alpha$ shifts to the right along the x-axis, and $R_c$ decreases. By connecting all the points of $t_{MRR}$ corresponding to different single-pass shifts in E-O case (i.e. different $a$), we obtain the distorted transmission circle of the E-O RPS, which can be described by parametric equations:

$$\begin{cases} x = \dfrac{(a^2+1)r - a(r^2+1)\cos(\phi)}{a^2 r^2 - 2ar\cos(\phi) + 1} \\ y = \dfrac{a(r^2-1)\sin(\phi)}{a^2 r^2 - 2ar\cos(\phi) + 1} \end{cases} \quad (5)$$

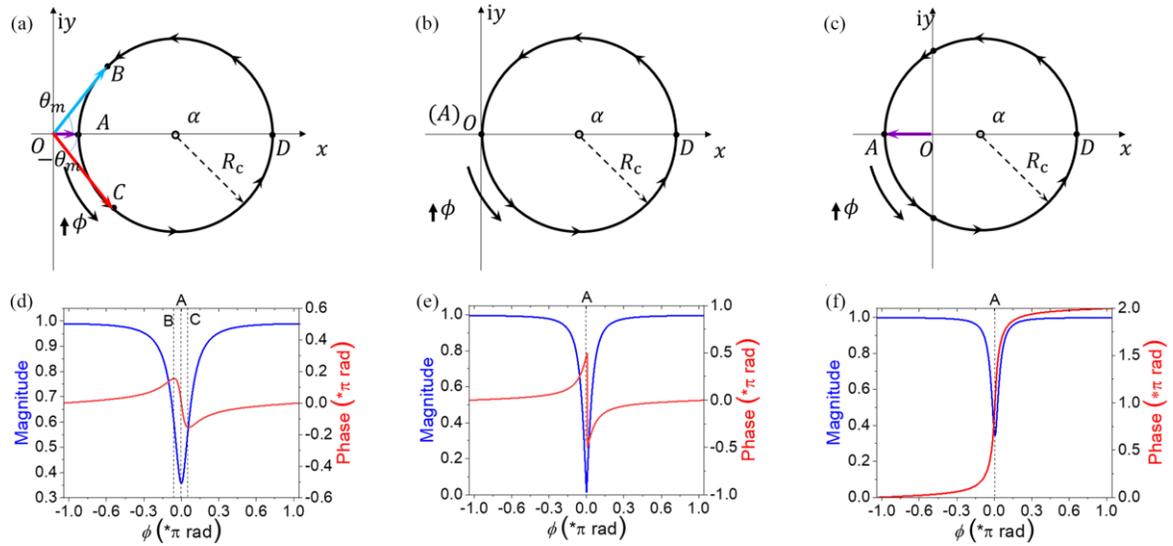

Fig. 3. Transmission circle chart when a RPS in (a) under-coupled regime, (b) critical-coupled regime and (c) over-coupled regime. Vector OA corresponds to $t_{MRR}$ in resonance. Vectors OB and OC are tangent to the under-coupled transmission circle chart with angle extremums of $\theta_m$ and $-\theta_m$ respectively. Magnitude transmission and effective phase shift curves when a RPS in (d) under-coupled regime, (e) critical-coupled regime and (f) over-coupled regime. Point A, B and C correspond to the vector OA, OB and OC in Fig. 3 (a)-(c).

Here, $a$ varies depending on several factors, including doping levels ($N$), bias voltages ($V$), and the RPS structure. Figure 4(b) depicts the distorted transmission circle of an E-O RPS according to APPENDIX A. If we examine the single-pass phase shift range from 0 to $2\pi$, there will be a disconnect, as shown in the magnified inset. This occurs because the phase shift by $2\pi$ does not return to the original point, due to changes in $a$ from carrier injection. An overlap point between the T-O transmission circle and the distorted one can be observed, as denoted in the inset, occurring when no bias voltage is applied to the ring. Additionally, it shows that when a large single-pass phase shift away from the resonance is induced, the transmission phase and magnitude closely resemble the original non-loss transmission circle. As depicted in Fig. 4(b), for the over-coupling regime, the maximum transmission difference $\Delta t_{MRR}$ corresponds to a merely insertion loss (IL) increase of

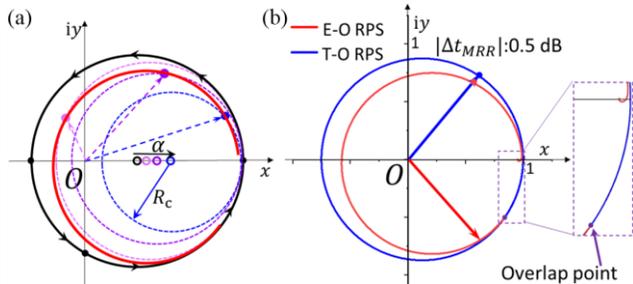

Fig. 4. (a)Schematic diagrams for distorted transmission circle trajectory formation of an EO RPS. Dashed circles: transmission circle with different $a$. (b) The transmission circle charts of T-O and E-O RPS.

~0.5 dB at $3\pi/4$ effective phase shift. This pattern holds true for all the three coupling regimes as when rings are driven away from their resonance, the coupling to the bus waveguide weakens significantly, resulting in decreased loss in the bus waveguide.

## III. OVER-COUPLED RPS BASED SWITCHING ELEMENTS

### A. Working Principle

Figure 5 illustrates the working principle of over-coupled RPS based switching elements (RPS-SEs). The structure is shown in Fig. 5(a). Multiple pairs of RPSs are embedded in the two arms of a balanced MZI. These RPSs operate in the over-coupled regime to provide a continuous phase shift from 0 to $2\pi$.

Starting with the most basic case, in which a single pair of RPS (RPS1 and RPS1') is utilized. To tackle asymmetrical transmission response between the bar and the cross state, a $\pi/2$ wideband phase delay is introduced in the upper arm. In this case, switching from the bar to the cross state can be accomplished by blueshifting the resonance wavelength of RPS1' to $\lambda_{0b}$, leading to a $3\pi/4$ phase advance at the operating wavelength ($\lambda_0$) in the lower arm, and redshifting the resonance wavelength of RPS1 to $\lambda_{0r}$, resulting in a $3\pi/4$ phase delay in the upper arm. Consequently, the RPS-induced phase difference amounts to $3\pi/2$. With the additional wideband phase delay of $\pi/2$, the phase difference ($\Delta\varphi$) becomes $2\pi$, allowing for switching from the bar to the cross state, as illustrated in Fig. 5(b). Conversely, interchanging the operation on RPS1 and RPS1', the RPS-induced phase difference becomes $-3\pi/2$. Adding an extra $\pi/2$ phase delay leads to $\Delta\varphi = -\pi$, enabling switch from the cross to the bar state. This push–pull control scheme effectively enables symmetrical switching, as depicted in Fig. 5(c). The dips on both sides of the passband are due to the resonance dip shifts of the ring under both blueshift and redshift conditions.

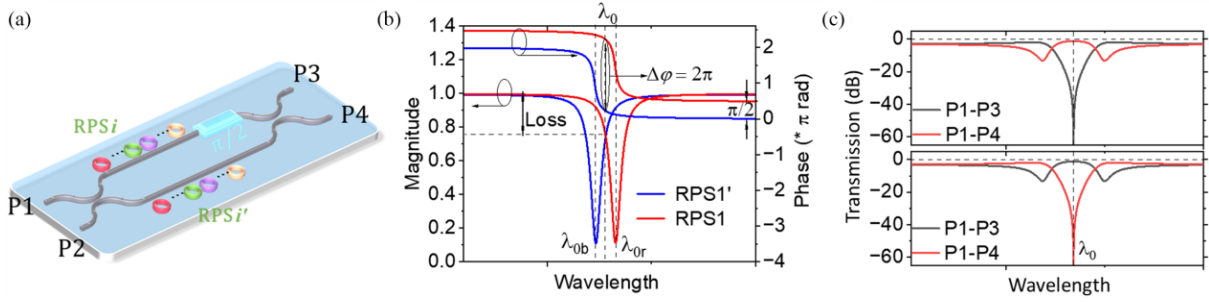

Fig. 5. Working principle of over-coupled RPS-SEs. (a) Schematic of an RPS-SE. (b) Power transmission and phase delay of RPS1 and RPS1' in cross state under pull-push configuration. (c) Power transmission of the over-coupled RPS-SE in cross state and bar state.

*B. Performance Analysis*

The performance metrics, including transmission, full width at half maximum (FWHM), 3dB-bandwidth (BW), robustness, extinction ratio (ER), and crosstalk, are analyzed utilizing the transmission circle chart presented above. The results are displayed in Fig. 6. In order to demonstrate the advantages of over-coupled RPS-SE, conventional critically coupled add-drop MRRs are used as counterparts.

The transmission circle chart is given in the inset of Fig. 6(a). For the T-O tuning case, vector OB and OC have a phase delay of $3\pi/4$ and a phase advance of $3\pi/4$ compared to OA, respectively, correspond to $t_{MRR}$ when the resonance wavelength of RPS1 redshifts to $\lambda_{0r}$ and the resonance wavelength of RPS1' blueshifts to $\lambda_{0b}$. The magnitude of OB and OC are the same. Since the transmission between the bar and cross state is symmetrical, their power transmission ($T_{bar}$, $T_{cross}$) can be treated equally and obtained simply through geometry:

$$T_{cross} = T_{bar} = (|OB| + |OC|)^2/4 \quad (6)$$

Detailed derivation of (6) and the corresponding resonance wavelength shifts ($\Delta\lambda_{0r}$, $\Delta\lambda_{0b}$) can be found in APPENDIX B. Contours of $T_{bar}$ in different $a$ and $r$ are also given in Fig. 6(a) (solid lines). Compared to the add-drop MRRs (dashed lines), the proposed SE exhibits an improved IL under the same coupling and single-pass loss conditions. This is because the operating point of RPSs deviate from its resonance wavelength by $\Delta\lambda_{0r}$.

For a precise analysis of the IL in E-O RPSs, the variation of $a$ has to be taken into consideration. According to Section II and APPENDIX A, for a specific RPS where $R$ and $N$ are determined, analyzing the impact of changes in $a$ during switching hinges on pinpointing the distorted transmission circle chart, specifically identifying the $a(R,V,N)$ when $\varphi$ is $3\pi/4$ and $-3\pi/4$, respectively. Although (6) remains valid, |OB| and |OC| are no longer equal. Instead, different values of $a$ corresponding to the $3\pi/4$ and $-3\pi/4$ phase shifts have to be used, as elucidated in APPENDIX B.

When $a$ and $r$ are determined, the BW and FWHM can be calculated as follows according to APPENDIX C:

$$\text{FWHM}(a,r) = \frac{\lambda_0^2 \Delta P_{1/2}(a,r)}{\pi n_g L} \quad (7)$$

$$\text{BW}(a,r) = \frac{c}{\lambda_0^2}\text{FWHM}(a,r) = \frac{c \cdot \Delta P_{1/2}(a,r)}{\pi n_g L} \quad (8)$$

where $\Delta P_{1/2}$ is the single-pass phase shift when the $T_{bar}$ declines to the half of its peak value. $c$ is the speed of light. Contours of BW in different $a$ and $r$ are given in Fig. 6(b) (solid lines). Compared to that of add-drop MRRs, the RPS-SEs demonstrate 3-4 times broadened BW under identical coupling and single-pass loss conditions. The BW broadening is especially pronounced under stronger coupling and higher single-pass loss. Such expanded BW results from the gradual phase change around the $\pm 3\pi/4$ operating point of the RPSs.

The operation of RPSs in the over-coupled regime and away from the resonance point also implies enhanced robustness and lower driving voltage resolution. Figure 6(c) (solid lines) illustrates the contours of the permissible maximum single-pass phase shift error ($\delta\phi$) due to finite voltage resolution for different $a$ and $r$ in the RPS-MZIs, necessary to achieve a 30 dB ER. This implies that it can endure three times the $\delta\phi$ compared to an add-drop MRR (dashed lines).

For random errors like fabrication variations and thermal fluctuations, we can assume that $\delta\phi$ follows a normal distribution N(0,$\sigma^2$). The stability can be characterized by the expected value of ER (E(ER)) within the 3$\sigma$ deviation. Figure 6(d) provides a quantitative comparison. $\Delta$ER is defined as the E(ER) difference of RPS-MZIs over add-drop MRRs. We set the superimposed contours of $\Delta$ER for different $a$ and self-coupling coefficients, with the standard deviation ranging from 0.005 to 0.03. This figure indicates that E(ER) values are almost equivalent in regions near critical coupling. However, in regions of over-coupling, as the difference between $a$ and self-coupling coefficient grows, the E(ER) of the RPS-SE surpasses that of the add-drop rings, indicating enhanced stability. The contours under different standard deviations, represented by lines of the same color, nearly coincide with each other except for a slight divergence in the upper right corner. This indicates that the superiority of RPS is maintained across various variances.

To ensure optimal performance of multi-wavelength switching with multiple pairs of RPSs, two additional criteria are: 1) the FSR of the RPSs should be large enough to accommodate all wavelength channels; 2) FWHM of the passband of SEs should be narrow enough to minimize crosstalk from adjacent wavelength channels. The first criterion can be achieved through the careful design of the FSR for each RPS. With respect to the second criterion, crosstalk from adjacent wavelength channels will lead to passband peak drift, reduced ER and additional IL [18]. To assess the impact of crosstalk, ER has been selected as the metric. According to APPENDIX D, the deterioration of ER is caused by the unbalanced ratio ($\alpha$) and the deviation of $\Delta\varphi$ from $-\pi$ or $2\pi$. As illustrated in Fig. 6(e), the channel $i$+1 will introduce different transmission power crosstalk ($\delta T_r$, $\delta T_b$) and phase crosstalk ($\delta P_r$, $\delta P_b$) on channel $i$ due to the different resonance wavelength of RPS $i$+1 and RPS $i$+1'. ($\delta P_r$, $\delta P_b$) and ($\delta T_r$, $\delta T_b$) are both determined by channel spacing $\Delta\lambda$ (assuming channels are equally spaced by $\Delta\lambda$) and the FWHM, as detailed in APPENDIX D. Therefore, achieving the desired ER requires meticulous control of the $\Delta\lambda$ and the FWHM. Considering crosstalk from all the other channels, the ER for channel i ($ER_i(\alpha_i, \Delta\varphi_i)$) can be calculated using transfer function method according to APPENDIX D. Figure. 6(f) presents ER contours varying with $a$ and $r$ across different total wavelength channel counts. A noticeable transition to higher ER values occurs as we move diagonally upward across the plot. This indicates that increasing the number of channels requires a lower single-pass loss and reduced coupling intensity to minimize crosstalk and achieve higher ER.

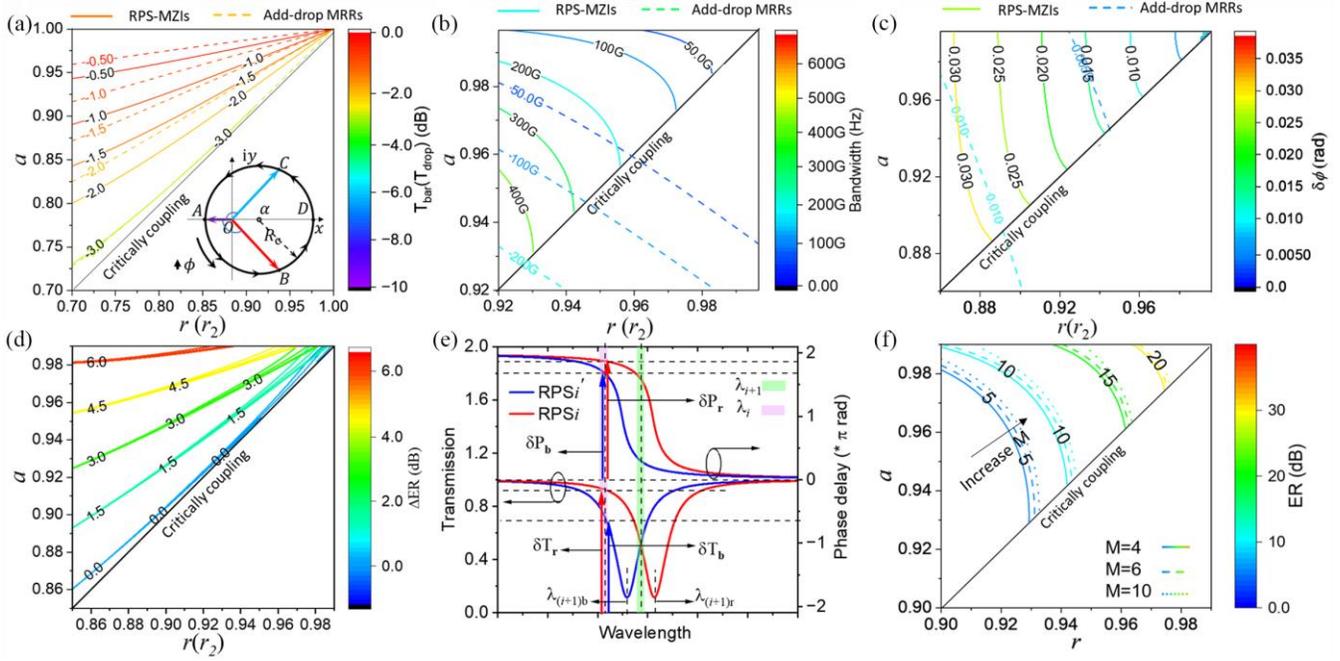

Fig. 6. Performance metrics of over-coupled RPS-SEs. (a) $T_{bar}$ Contours of the proposed SE in different $a$ and $r$ (solid lines). $T_{drop}$ Contours of critically coupled add-drop MRRs in different $a$ and self-coupling coefficient ($r_2$) (dashed lines). Inset: Transmission circle chart under the pull-push configuration. (b) Bandwidth contours of the proposed SE in different $a$ and $r$ (solid lines) and critically coupled add-drop MRRs in different $a$ and $r_2$ (dashed lines). (c) Permissible maximum single-pass phase shift error ($\delta\phi$) contours to achieve 30 dB ER for the proposed SEs in different $a$ and $r$ (solid lines) and the critically coupled add-drop MRRs in different $a$ and $r_2$ (dashed lines). (d) Superimposed expected ER difference contours ($\Delta$ER) of the proposed SE over critically coupled add-drop MRRs in different $a$ and $r(r_2)$ with the standard deviation ($\sigma$) ranging from 0.005 to 0.03. (e) Schematic diagram of crosstalk from adjacent channels. (f) ER contours in varying $a$ and $r$ with total wavelength channel count $M$=4,6 and 10.

## IV. UNDER-COUPLED RPS BASED SWITCHING ELEMENTS

While over-coupled RPS-SEs exhibit many advantages, they encounter challenges when using smaller RPSs with wide FSR for wideband operation. Smaller RPSs lead to higher bending loss (i.e., smaller $a$) and limited coupling between the ring and bus waveguide (i.e., higher $r$), making it challenging to reach the over-coupling regime ($a > r$).

To address this challenge, under-coupled RPSs stand out as a promising solution because they only require $r > a$, which eases the coupling constraints. As discussed in section II, there are two extremums ($\pm\theta_m$) on the effective phase shift curve, corresponding to the tangent vectors OB and OC. Therefore, in a push-pull configuration, the phase change range spans from $-2\theta_m \sim +2\theta_m$. This limited phase range corresponds to the upper limit of $r$, which can be determined by solving the simultaneous equations of the transmission circle and its tangent vector. Its expression is as follows:

$$a < r < \frac{\sqrt{2}}{2}\sqrt{\frac{a^4k^2 + k(k-\Omega) + a^2(2+k\Omega)}{a^2(1+k^2)}} \quad (9)$$

where:

$$k = \tan(\theta_m)$$
$$\Omega = \sqrt{k^2 + a^4k^2 + 2a^2(2+k^2)}$$

For switching application, a phase change of $\pi$ is required, so $\theta_m = \pi/4$. The corresponding transmission chart is presented by the inset in Fig. 7. When $r$ does not equal its upper limit, there are four vectors with angles of $\pm\pi/4$ intersecting transmission circle (OB, OC, OE, OF). The two vectors with a larger amplitude (OB, OC) are chosen to minimize transmission loss. Contours of the transmission are depicted in Fig. 7 according to (6). As $r$ increases to its upper limit, the transmission loss shows a gradual increase from approximately 3dB to around 6dB. Therefore, the decision to employ under-coupled RPSs involves weighing these trade-offs: broader spectral flexibility and more channel capacity against higher losses and reduced phase control. Likewise, the single-pass phase shift ($\Delta P_{0r}$, $\Delta P_{0b}$) and their corresponding resonance wavelength shifts ($\Delta_{\lambda_{or}}$, $\Delta_{\lambda_{ob}}$), FWHM, BW and ER can be calculated according to APPENDIX B, C, D.

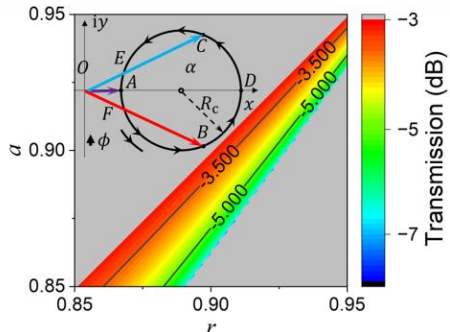

Fig. 7. Contours of $T_{bar}$ in different $a$ and $r$. Inset: under coupled transmission circle chart under pull-push configuration.

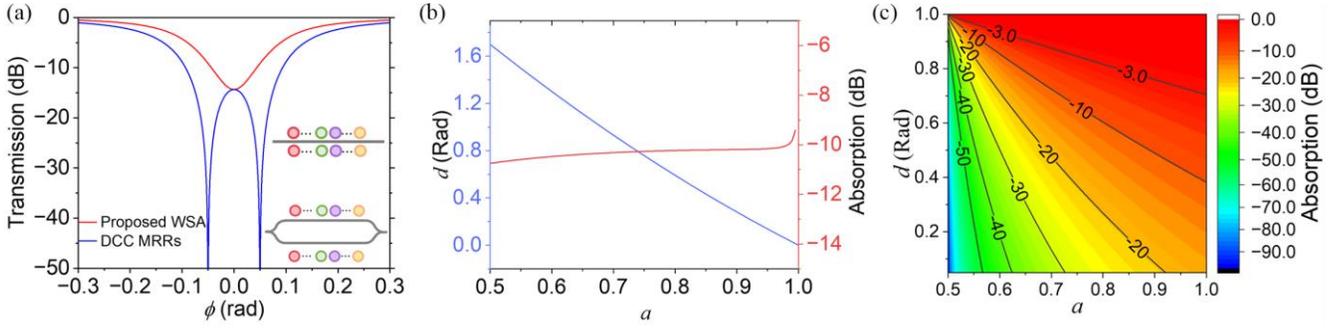

Fig.8. Performance metrics of WSA (a) Power transmission demonstrating the performance of the proposed WSA and DCC MRRs when the resonance wavelength of critically coupled MRRs has a deviation from the operational wavelength. (b) The critical condition for the absence of peaks within the rejection band (blue line) and the corresponding absorption values (red line). (c) Absorption contours of the proposed WSA in different deviations $d$ and $a$.

## V. CRITICALLY-COUPLED RPS BASED ABSORBERS

As depicted in Fig. 3(b) and (e), a phase singularity emerges at resonance when the RPS operates in the critical coupling regime. This singularity, associated with zero transmission, can be harnessed by wavelength-selective absorber (WSA) [19]. Previous work proposed singly coupled critical MRRs (SCC MRRs) as absorbers [20]. However, SCC MRRs are susceptible to fabrication discrepancies and environmental interferences; even slight deviations can significantly reduce absorption efficiency. To mitigate this, including supplementary critically coupled rings can offer some relief. Nonetheless, dual coupled critical MRRs (DCC MRRs) tend to broaden the rejection bandwidth, and even slight inconsistencies in the resonance wavelengths of the two rings introduce a peak between the two discrepant resonance points, as shown in Fig. 8(a), leading to undesired signal leakage and data integrity loss. The bottom inset of Fig. 8(a) gives our proposed design: pairs of RPSs affiliated with the targeted absorption channels in the RPS-MZI are designed to operate in the critical coupling regime. This configuration transforms the critically coupled RPS-MZI into an efficient WSA, capable of achieving a narrow yet uncompromised rejection bandwidth. Within the rejection band, the proposed WSA maintains a flattened rejection band even with deviations in the two rings' resonance, as depicted in Fig. 8(a). By sweeping the phase deviating from resonance ($d$) and $a$, we can find the critical condition to suppress trough splitting, as depicted in Fig. 8(b), implying that absorption below 10 dB is associated with the emergence of a new peak within the stopband.

In addition to its inherent stability, the intentional resonance wavelength shift of the MRRs allows for the WSA to reconfigure its rejection bandwidth, while maintaining a commendable absorption efficiency. The contours of absorption in different single-pass phase shift deviation $d$ and $a$ are given in Fig. 8(c). The region below -20 dB is deemed appropriate for absorber design as it represents the absorption ≥99%. The corresponding bandwidth at the -20 dB rejection ($BW_{-20\text{dB}}(a,r)$) can be calculated according to APPENDIX C, where $\beta = 0.01$.

To enable multichannel absorption, a certain range should be intentionally reserved within the FSR spectrum region. When all channels need to be set to ON, the resonance wavelengths of all MRRs will be set to the same $\lambda_{ON}$ within the reserved spectrum region. Similarly, the FSR requirement and crosstalk can be analyzed following the methods in Section III.

## VI. ULTRA-DYNAMIC MULTI-DIMENSIONALE SWITCHING FABRIC

RPS-based SEs and WSAs enable efficient independent control over space and wavelength at nanosecond speeds. Using various TDM algorithms [21], we can achieve a highly dynamic optical switching fabric that effectively utilizes the dimensions of time, space, and wavelength. A notable application of this is its support for ultra-dynamic optical burst switching, which allocates packet headers and main payloads to different wavelength channels and directs them to different space ports. In scenarios with extremely high data flow, packets can be further interleaved over time to share a common space or wavelength channel. This significantly enhances bandwidth efficiency, reduces idle times, increases overall throughput, and offers the flexibility to reroute traffic instantly in case of link failures or congestion.

Leveraging the novel analytical tool proposed above, a multi-dimensional, ultra-dynamic switch fabric is presented with an example port count of 8×8×8λ, presenting a general design methodology. The selection of topology plays a critical role in determining the performance of a switch fabric. Commonly utilized topologies can be divided into 2 categories: broadcast & select (B&S) and route & select (R&S) [22]. The B&S topology divides input data streams to all conceivable destinations and applies selective absorption except for intended recipients, which is suitable to use WSAs as SEs. It requires $N^2$ WSAs for an N×N switch radix. In contrast, the R&S topology accentuates network efficiency by channeling data solely to its intended endpoint. This approach affords a heightened level of control over network resources. Dilated Banyan topology stands out as a great host to the RPS-SEs considering its complete immunity to first-order spatial crosstalk. An N×N dilated Banyan switch requires $2N(N-1)$ SEs in total [23].

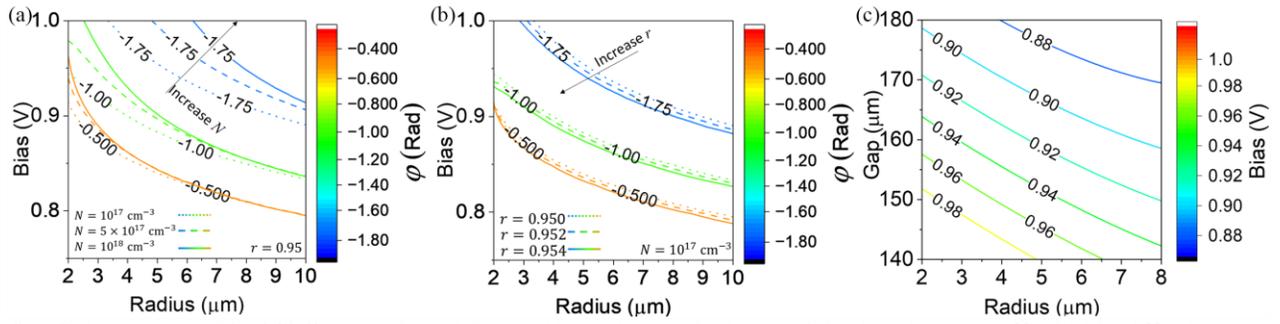

Fig. 9. E-O characteristics of the RPS. Contours of $\varphi$ in different (a) bias voltages and radiuses with fixed $r$ and increasing $N$ and (b) fixed $N$ and increasing $r$. (c) Contours of required bias voltage for red shift state ($V_r$)) in different radius and gap.

### A. Circuit-level specifications break-down

The 8 wavelength channels are assigned by the Continuous-Wave Wavelength Division Multiplexing Multi-Source Agreement (CW-WDM MSA)[24]. The required FSR is approximately 9 nm so over-coupled RPS-SEs are deemed suitable. Therefore, to achieve the overall performance of $ER_{fabric} \geq 17$ dB, $BW_{fabric} \geq 30$ GHz and $IL_{fabric} \leq 6$ dB, the corresponding performance for each SE can be estimated by[25]:

$$IL_{SE} \leq \frac{IL_{fabric}}{2\log_2 m} \text{ (dB)}$$
$$ER_{SE} \geq ER_{fabric} + 10\log_{10}(2\log_2 m) \quad (10)$$
$$BW_{SE} \geq \frac{c \cdot \Delta P_\beta(a,r)}{\pi n_g L}$$

where $m$, $IL_{SE}$, $ER_{SE}$, $BW_{SE}$ are the stage number of the fabric, IL, ER and BW of each SE. When N=8, 6 SE stages are used in the dilated Banyan switch, so the break-down metrics for each SE are $IL_{SE} \leq 1\ dB$, $ER_{SE} \geq 25\ dB$, $BW_{SE} \geq 50 GHz$.

### B. Switch element design space

To attain a lower $r$ while preserving a compact footprint, a racetrack-type RPS with a 1.5 µm-long straight coupled waveguide is used. The injection type of E-O racetrack RPS has the following parameters: $W$: 380 nm, $H$:220 nm, $H_{slab}$: 62.2 nm, $W_i$:0.5 µm, $W_p$: 1.3 µm, $W_n$: 1.3 µm. The corresponding simulated E-O performance is given in Fig. 9. The inherent resonant wavelength of the RPS is set at $\lambda_{ib}$ and a bias can be applied to redshift it to $\lambda_{ir}$. Contours of $\varphi(\phi(R,V,N,r))$ for different $R$ and $V$ values and with changing $N$ and $r$ are given in Fig.9(a) and (b), respectively, according to APPENDIX B. As $N$, the doping concentration, increases, a higher $V$ is required to reach the same phase at a given R. This increased voltage is necessary to counteract the built-in electric field that results from high doping levels. However, an increase in $r$ leads to a reduction in the required $V$ for a given $R$ to achieve the same phase. This occurs because a larger $r$ implies a weaker coupling strength, which results in a steeper phase change curve, thereby enhancing the efficiency of phase tuning. $N = 10^{17} cm^{-3}$ is selected as the doping concentration for efficient phase tuning. Following APPENDIX B, $a$ and the corresponding bias voltage for redshift ($V_r$) can be determined, as shown in Fig. 9(c). Since $V_b$ and $V_r$ has been determined, the design space for the RPS is depicted in Fig. 10 by finding the intersection of all required performance boundaries according to Section III and translate ($a, r$) into (Radius, Gap) according to [26]. Within the design space, the requisite performance is consistently achievable, despite variations in $a$ during the tuning process.

### C. Validation through FDTD Simulations

According to the final design space outlined, we choose R=5.5 um, G=190 nm, and $N = 10^{17}$ cm$^{-3}$ as the optimal set of parameters to construct the multi-dimensional 8×8×8λ switch fabric. This is also the design that we decide on running the FDTD simulations for validation. The fourth wavelength channel exhibits the worst performance with $L_{SE}$, $ER_{SE}$ and $BW_{SE}$ being 0.93 dB, 25.2 dB, and 54.92 GHz as shown in Fig. 10(b), indicating the worst crosstalk. A comparison with our analytical values (0.77 dB, 25.56 dB, and 55.43 GHz) reveals a good agreement with an acceptable level of error. The deviation can be attributed to the linear dispersion approximation employed in the analytical method and potential inaccuracies inherent in the FDTD simulation method. Nonetheless, it is worth noting that all targeted metrices are consistently met, even in the worst-case scenario. This affirms the robustness of our design methodology, particularly for expedited prototyping.

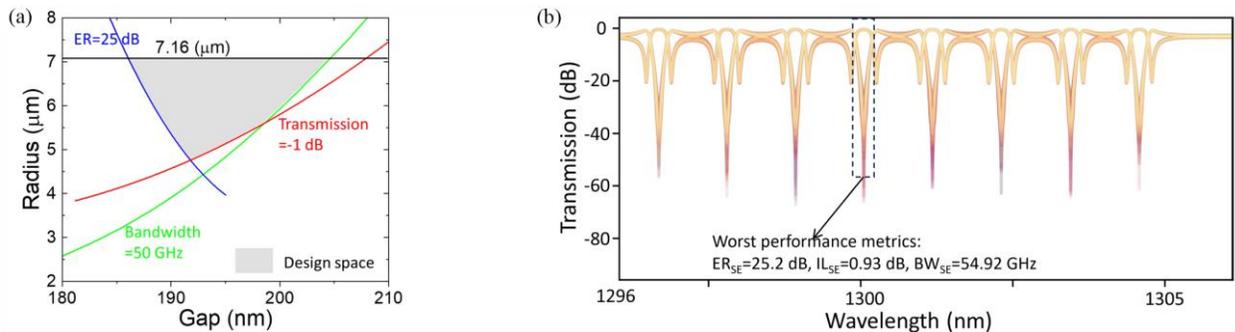

Fig. 10. (a) Design spaces for the EO-RPS. (b) Superimposed simulated transmission spectrum for all 256 switching states.

## VII. CONCLUSION

By gaining insights into the magnitude and phase response of RPSs with an innovative compact model, we present comprehensive analyses for RPS-SEs and show its power in manipulating signal in spatial, spectral, and temporal domains. Integrating such RPSs with MZIs not only greatly enlarges its design space for efficient spectral multiplexing but also enables E-O nanosecond-scale phase tuning that was sparsely achieved with conventional add-drop MRRs. Further leveraging such a powerful analytical tool, we detail the superiority of RPS-MZIs over add-drop MRRs on metrices such as loss, extinction ratio, bandwidth, robustness and driving voltage. We also show the design robustness in under- and over-coupled regimes for wavelength routing, and in critical coupling condition for multi-band absorbing. The effectiveness of the proposed methodology is confirmed through rigorous FDTD simulations, establishing it as a potent tool for efficient and rapid prototype design and potentially making a change in the field of optical switching and signal processing.

## APPENDIX A: DEPENDENCIES OF $a$ AND $\varphi$

Complete operating characteristics of E-O RPS can be determined by the power attenuation coefficient, ($\alpha_{EO}(N,V)$) [dB/cm]), and the variation in effective refractive index ($\Delta n_{\text{eff}}(N,V)$) due to the plasma dispersion effect. Both factors are a function of the applied bias voltage ($V$) and the doping level ($N$). Assuming the p-type and n-type doping level are identical and considering both carrier injection and curvature of waveguide, $a$ can be expressed as:

$$a(R,V,N) = a_{bend} a_{EO} \quad (A1)$$

where:

$$a_{bend} = 10^{-(\frac{\alpha_{bend}}{20}) \cdot \frac{2\pi R}{10000}}$$

$$a_{EO} = 10^{-(\frac{\alpha_{EO}}{20}) \cdot \frac{(2\pi R + 2Lc)}{10000}}$$

The power attenuation coefficient due to waveguide curvature, denoted as $\alpha_{bend}$ [dB/cm], correlates with the radius of waveguide curvature ($R$) and adheres to a power law relationship [12]. The single-pass phase shift can be given by:

$$\phi(R,V,N) = 2\pi(2\pi R + 2Lc)\Delta n_{\text{eff}}(N,V)/\lambda_i + \phi_0 \quad (A2)$$

where $\phi_0$ is the single-pass phase shift when no bias voltage is applied. (A1) and (A2) imply that $a$ can be expressed as a function of $\phi$, providing insight into how $a$ changes with varying values of $\phi$. After conducting simulation and fitting, we establish the relationship between a and $\phi$ : $a(\phi) = 0.9964\exp(-((\phi(R,V,N) - \phi_0 - 0.1802)/3.8292)^2)$ under a specific design parameters: W: 380 nm, H: 220 nm, $H_{\text{slab}}$: 62.2 nm, Wi: 0.5 μm, Wp: 1.3 μm, Wn: 1.3 μm, N: $10^{17}$ /cm$^{-3}$ The transmission circle charts are given in Fig. 4(b) according to (5), clearly depicting the distortion and contraction under the influence of plasma dispersion effect. Here we set $\phi_0 = 0.15$ so that with the increase of V, the entire single-pass shift changes from a positive initial value, through zero, to a negative value. This adjustment facilitates setting the redshift states (negative single-pass shift) and the blueshift (positive single-pass shift) states by moving ring's resonant peak unidirectionally, without tuning across the entire FSR and higher electro-absorption loss.

## APPENDIX B: POWER TRANSMISSION FOR BAR AND CROSS STATE

Power transmission coefficients for P3 and P4 of a MZI can be given, respectively, by:

$$T_3(\varphi_u, \varphi_l, |t_u|, |t_l|) = |t_u|^2/4 + |t_l|^2/4 - |t_u||t_l|\cos(\varphi_u - \varphi_l)/2 \quad (B1)$$

$$T_4(\varphi_u, \varphi_l, |t_u|, |t_l|) = |t_u|^2/4 + |t_l|^2/4 + |t_u||t_l|\cos(\varphi_u - \varphi_l)/2 \quad (B2)$$

where $|t_u|(|t_l|)$ and $\varphi_u(\varphi_l)$ are the optical field amplitude transmission coefficient and the corresponding phase on the upper arm (lower arm), respectively. For T-O transmission circle, as shown in the inset of Fig. 6(a), the power transmission of single RPS can be calculated by simple geometry:

$$|t_u|^2 = |t_l|^2 = |OB|^2 = |OC|^2$$
$$= \frac{1}{2}\left(\frac{r - a^2 r}{1 - a^2 r^2}\right.$$
$$\left. + \sqrt{\frac{-2a^2 r^2 + 2a^2 - r^2 + 2a^2 r^4 - a^4 r^2}{(1 - a^2 r^2)^2}}\right)^2 \quad (B3)$$

By substituting (B3) and the corresponding phases into (B1), we have:

$$T_{cross} = T_{bar} = (|OB| + |OC|)^2/4 = |OB|^2 = |OC|^2 \quad (B4)$$

By solving (B4) and (3) simultaneously, the single-pass phase shifts ($\Delta P_{0r}, \Delta P_{0b}$) and their corresponding resonance wavelength shifts ($\Delta_{\lambda_{0r}}, \Delta_{\lambda_{0b}}$), can be given, respectively, as:

$$\Delta P_{0r} = -\Delta P_{0b}$$
$$= arccos\left(\frac{-a^2(a^2+3)r^3 + 3a^2r + r + (a^4r^4 - 1)\Psi}{2a(-a^2r^4 + 1 + (a^2r^3 - r)\Psi)}\right) \quad (B5)$$

$$\Delta\lambda_{0r} = -\Delta\lambda_{0b}$$
$$= \text{FSR} \cdot \frac{P_{0r}}{2\pi} = \frac{\lambda_0^2}{2\pi n_g L}\Delta P_{0r} \quad (B6)$$

where:

$$\Psi = \sqrt{\frac{2a^2r^4 - (a^2+1)^2 r^2 + 2a^2}{(a^2r^2 - 1)^2}}$$

For E-O transmission circle, (B3), (B4), (B5), and (B6) persist in their validity. However, a crucial adjustment is required: replace the $a$ with a distinct value for the calculation of |OB| and |OC|, which corresponds to the $3\pi/4$ and $-3\pi/4$ phase shifts. For a given RPS, The effective phase shift can be given by:

$$\varphi(\phi(R,V,N,r)) = \varphi((2\pi R + 2Lc)\frac{2\pi}{\lambda_i}\Delta n_{\text{eff}}(N,V - V_b) + \Delta P_{0b}(a(R,V_b,N),r)) \quad (B7)$$

where $V_b$ is the applied bias when RPS in blueshift state and can be deliberately predetermined. The applied bias for the redshifts ($V_r$) can be obtained by letting (B7) equal $-7\pi/4$.

Since we have already obtained $V_b$ and $V_r$, $a$ corresponding to the $3\pi/4$ and $-3\pi/4$ phase shifts can be determined according to (A1). Consequently, $|OB|$ and $|OC|$ can be determined by substitute $a(R, V_b, N)$ and $a(R, V_r, N)$ into (B3) respectively. Then the power transmission of $T_{bar}$ and $T_{cross}$ can be given by (6).

## APPENDIX C: 3DB BANDWIDTH AND FWHM

The deviation of the operating wavelength from $\lambda_i$ will lead to the decrease of transmission according to (B1) and (B2). When the power transmission declines, we have:

$$T_3(\varphi(\Delta P_{0r} + \Delta P_\beta), \varphi(\Delta P_{0b} + \Delta P_\beta), |t_{MRR}(\Delta P_{0r} + \Delta P_\beta)|, |t_{MRR}(\Delta P_{0b} + \Delta P_\beta)|) \quad (C1)$$
$$= \beta \cdot T_{bar}$$

where $\Delta P_\beta$ is the additional single-pass phase shift caused by the operating wavelength deviating from $\lambda_i$. $\Delta P_\beta$ can be obtained by solving (1) (B1), (B5) and (C1) simultaneously. When $\beta = 1/2$, $\Delta P_\beta$ corresponds to the FWHM where the transmission drops to half of its peak value. Under the assumption of linear dispersion approximation, the FWHM and BW can be given by (7) and (8). When $m$ the RPS-SEs cascade, the BW and FWHM of the whole system can be obtained by let $\beta = (1/2)^{1/m}$.

## APPENDIX D: DERIVATION OF EXTINCTION RATIO

If we define the unbalanced ratio ($\alpha$) for the two arms of MZI as:

$$\alpha = |t_u|/|t_b| \quad (D1)$$

Then ER can be expressed as:

$$ER(\alpha, \Delta\varphi) = \frac{T_3}{T_4} = 1 - \frac{4\cos(\Delta\varphi)}{\alpha + \frac{1}{\alpha} + 2\cos(\Delta\varphi)} \quad (D2)$$

Therefore, the $ER$ for channel i ($ER_i(\alpha_i, \Delta\varphi_i)$) under crosstalk from all the other channels can be given by:

$$ER_i(\alpha_i, \Delta\varphi_i) = \left(1 - \frac{4\cos(\Delta\varphi_i)}{\alpha_i + \frac{1}{\alpha_i} + 2\cos(\Delta\varphi_i)}\right) \cdot T_{bar} \quad (D3)$$

where:

$$\Delta\varphi_i = \sum_{\substack{k=1 \\ i \neq k}}^{M-1} \left(\varphi(\Delta P_{kb} + \Delta S_k) - \varphi(\Delta P_{kr} + \Delta S_k)\right) + \pi$$

$$\alpha_i = \prod_{\substack{k=1 \\ i \neq k}}^{M-1} \frac{|t_{MRR}(\Delta P_{kb} + \Delta S_k)|}{|t_{MRR}(\Delta P_{kr} + \Delta S_k)|} \quad (D4)$$

$$\Delta P_{kb} = \Delta P_{kr} = 2\pi \cdot \Delta_{\lambda_{kr}}/FSR$$

Here, M is the total wavelength channel count. $\Delta P_{kb}$ and $\Delta P_{kr}$ are the single-pass phase shift of the $k$-th pair of RPSs corresponding to the redshift wavelength $\Delta\lambda_{kr}$, blueshift wavelength $\Delta\lambda_{kb}$. $\Delta S_k$ is the single-pass phase shift of the $k$-th pair of MRPSs when operating wavelength deviating from $\lambda_k$ to $\lambda_i$.